\documentclass[12pt,twocolumn]{iopart}
\usepackage{iopams}
\usepackage{hyperref}
\usepackage{bbm}
\usepackage{amsfonts, amssymb}
\usepackage{color}
\usepackage{graphicx,epsfig}
\usepackage{pifont} 
\usepackage{subfigure}

\begin{document} 
\title{Quantum entanglement of anharmonic oscillators}
\author{Chaitanya Joshi$^1$, \footnote{Author to whom any correspondence should be addressed.}, Mats  Jonson$^{1,2,3}$, 
Erika  Andersson$^1$ and Patrik  \"Ohberg$^1$} 
\address{$^1$ SUPA, Department of Physics, Heriot-Watt University, Edinburgh, EH14 4AS, UK}
\address{$^2$ Department of Physics, University of Gothenburg, SE-412 96 G{\"o}teborg, Sweden} 
\address{$^3$ Department of Physics, Division of Quantum Phases \& Devices, Konkuk University, Seoul 143-701, Korea}
\ead{cj53@hw.ac.uk} 
\begin{abstract} 
We investigate the quantum  entanglement dynamics  of undriven anharmonic (nonlinear) oscillators  with
quartic potentials. We first consider the indirect interaction  between two such nonlinear oscillators mediated by a third,
linear oscillator and show that it  leads to a time-varying entanglement of the oscillators, the entanglement being
strongly influenced by the nonlinear oscillator dynamics.  In the presence of dissipation, the role of nonlinearity is strongly manifested in the steady state dynamics of the indirectly coupled anharmonic oscillators.
We further illustrate the effect of nonlinearities by studying the coupling between 
an electromagnetic field in a cavity with one  movable mirror which is modeled as a nonlinear oscillator. 
For this case we present a full analytical treatment, which is valid in 
a regime where both  the nonlinearity and the coupling due to radiation pressure is weak. 
We show that, without the need of any conditional measurements on the cavity field, the state of the movable mirror 
is non-classical 
as a result of the combined effect of the intrinsic nonlinearity and the radiation pressure coupling.
This interaction is also shown to be responsible for squeezing the movable mirror's position quadrature beyond the minimum uncertainty state even when the mirror is initially prepared in its ground state. 
\end{abstract} 
\pacs{1315, 9440T} 
\submitto{\JPB} 
\maketitle
\section{Introduction}
\label{sec:introdu}
\noindent The last decade has seen a surge of interest in investigating the quantum properties of micro- or nanomechanical systems \cite{schwab}. There have been  numerous efforts to prepare entangled or other non-classical states of such mechanical systems. With the advent of techniques such as laser cooling~\cite{iwra}, these mechanical systems  can be  cooled sufficiently close to their ground states and hence tailored to mimic the physics of  quantum harmonic oscillators to a very good approximation. There  are many proposals which aim to investigate the dynamics of such mechanical systems operating deep in the quantum regime. These include  coupling them to other quantum 
systems  such as an ultracold atomic Bose-Einstein Condensate (BEC)~\cite{ptre,davi}, a Cooper pair box~\cite{adar,bose} and even entangling two distant oscillators~\cite{adar,bose,stef,josh}. Often the main motivation is to test the foundations of quantum mechanics, but these nano-mechanical systems also turn out  to be extremely good candidates for applications such as ultra-sensitive measurement devices~\cite{kjva}.

  The  physics of anharmonic oscillators has been studied in great detail by several authors both in the classical and quantum domain \cite{kkle,mil2,lcho,vpea}. In particular, Milburn has invesigated the quantum and classical dynamics of an anharmonic oscillator in phase space \cite{mil2} and has shown that decoherence induced state reduction results in quantum to classical crossover in a nonlinear oscillator. Also in \cite{vpea} a quantum master equation has been derived for a doubly clamped driven nonlinear beam.
  
Many of the existing schemes, which aim at exploring the quantum dynamics of 
nano-mechanical systems, treat them as {\em harmonic} oscillators. However, an {\em anharmonic} (nonlinear) oscillator in the quantum regime offers a number of intriguing new possibilities for quantum state preparation and manipulation. One 
of many motivations for 
studying nonlinear oscillators is
that by active cooling techniques, such as laser cooling, the thermal fluctuations of  these nanomechanical systems can only be 
reduced to the standard quantum limit. If a reduction in noise is sought beyond 
this limit, then squeezing the quadratures of these mechanical oscillators is required. For this one typically relies on nonlinearities. There already exists many feasible schemes that explore the possibility of  squeezing the state of a mechanical oscillator \cite{prabl,ajij,ralmo}.  Moreover, coherent nonlinear effects are of great interest as they turn out to be important resources for processing universal quantum information with continuous variables \cite{slbr}.

 In the present work, however, we shall concentrate on the influence of an intrinsic nonlinearity on the entangled states of two such indirectly coupled oscillators.  We address a situation where the anharmonic oscillator is coupled to a second quantum system. Firstly we investigate the quantum dynamics of two such anharmonic oscillators interacting with a linear oscillator. We show that as a result of 
indirect interactions mediated by the linear oscillator, the two nonlinear oscillators exhibit a time-varying entanglement. Interestingly, we find that the effect of nonlinearity is much more pronounced for certain initial states. When dissipation is included, the effect of nonlinearity strongly governs the steady state evolution of the indirectly coupled nonlinear oscillators.

As a second illustration of the effect of the nonlinearities we investigate the unitary evolution of a cavity mode interacting with a movable mirror which is  modeled as an anharmonic oscillator.  We provide a full analytical treatment  of a physical model that describes  this interaction  in a regime where both  the nonlinearity and the coupling due to radiation pressure is weak. We show that  unitary evolution  results in time-dependent entanglement between the oscillator and the cavity mode. Moreover, under the joint action of radiation pressure coupling and intrinsic nonlinearity the movable mirror will also exhibit non-classical dynamics \cite{jonas}.

Nonlinear effects are typically small in nano-cantilevers, since the amplitude of 
their oscillations are inevitably small compared to their
length.
Moreover, it is difficult to control the nonlinearities externally. In this paper we propose to use an electromagnetic setup based on a Helmholz coil configuration, 
where the nonlinearity stems from the fact that the energy due to the interaction between the magnetic field produced by the coils and permanent magnets at the tips of the cantilevers has a term that depends on the fourth power of the deflection of a tip from its equilibrium position.  This allows us to externally tune the strength of the nonlinearity, which may be difficult in other realisations \cite{ajij,ralmo}. 

The paper is organized as follows. In section \ref{sec:model} we introduce the theoretical model describing, in detail, the indirect interaction of two anharmonic oscillators mediated via a linear oscillator. It is followed by section \ref{sec:cavity}, in which we investigate  the coherent interaction  between an anharmonic oscillator and a  quantized cavity mode. In section \ref{sec:out} we briefly sketch a scheme to induce nonlinearities for the  mechanical oscillator, and finally we present our
conclusions  in section \ref{sec:concl}.

\section{Indirectly coupled anharmonic oscillators}
\label{sec:model}
\noindent 
In this section
we shall explore the quantum dynamics of two anharmonic oscillators, which interact
with 
the same linear oscillator. We will keep the theoretical treatment general at this point, but 
in section \ref{sec:out} we will discuss a potential realisation of the required nonlinearites.

\subsection{Unitary Dynamics}
\noindent
Consider two identical  micro- or nanomechanical oscillators each of mass $m$ and operating in the quantum regime with  fundamental vibrational frequency $\omega_{m}$. Denoting 
the position and momentum operators of each oscillator by  $\hat{q}_{i}$ and $\hat{p}_{i}$, where $i=1,2$, the 
free evolution of the oscillators 
is governed by the Hamiltonian
\begin{equation}
\hat{H}=  \sum _{i=1} ^{2} \left [\frac{\hat{p_{i}}^{2}}{2m}+ \frac{m \omega_{m}^{2}{\hat{q_{i}}}^{2}}{2} \right ].
\end{equation}
If we can modulate the potential seen  by the oscillator such that 
there is an additional  term proportional to $\hat{q}_{i}^{4}$, then this will introduce an effective nonlinearity for the mechanical oscillator. The Hamiltonian of two such independent  anharmonic oscillators then takes the form
\begin{equation}\label{hamclass}
\hat{H}_{1}=  \sum _{i=1} ^{2}\left [\frac{\hat{p_{i}}^{2}}{2m}+ \frac{m \omega_{m}^{2}{\hat{q_{i}}}^{2}}{2}+\tilde{\beta} \hat{q}_{i}^{4} \right ],
\end{equation}
where $\tilde{\beta} \hat{q}_{i}^{4} $ is the nonlinear interaction energy. Expressing the position and momentum operators of each oscillator as
\begin{eqnarray*}
\hat{q}_{1}=\sqrt{\frac{\hbar}{2 m \omega_{m}}}(\hat{a}^{\dagger}+\hat{a});~~~~~~  \hat{p}_{1}=i\sqrt{\frac{\hbar m \omega_{m}}{2 }}(\hat{a}^{\dagger}-\hat{a})\\
\hat{q}_{2}=\sqrt{\frac{\hbar}{2 m \omega_{m}}}(\hat{b}^{\dagger}+\hat{b}); ~~~~~~\hat{p}_{2}=i\sqrt{\frac{\hbar m \omega_{m}}{2 }}(\hat{b}^{\dagger}-\hat{b}),
\label{nayaeqn}
\end{eqnarray*}
where $\hat{a}^{\dagger}(\hat a)$ and $\hat{b}^{\dagger}(\hat b)$  are the creation (annihilation) operators for the vibron excitations of the two anharmonic  oscillators, and neglecting all the counter-rotating terms,  \eref{hamclass} takes
the form
\begin{equation}\label{hamnonline11} 
{\widetilde{\hat{H}}}/{\hbar} \approx  \omega_{m}( \hat{a}^{\dagger}\hat{a}+\hat{b}^{\dagger}\hat{b})
+\beta(\hat{n}_{a}^{2}+\hat{n}_{a})+\beta(\hat{n}_{b}^{2}+\hat{n}_{b}), 
\end{equation}
 where $\hat{n}_{a}$ and $\hat{n}_{b}$ are  the number operators of the two anharmonic oscillators  and $\beta$ is the nonlinearity strength. It is worth stressing that in general, for a driven  nonlinear oscillator, the oscillation frequency depends on the driving amplitude \cite{nonlinbe}, although  a single resonance frequency  can still be a valid approximation in the case of a very weak driving force. Moreover, in the present work we are considering a system of two undriven nonlinear oscillators, for which assigning a single resonance frequency seems to be  a reasonable assumption. Nonetheless, depending on the initial excitation amplitude, the nonlinear oscillator might exhibit multistable behavior. But as long as the initial average number of excitations  $\langle \hat{n} \rangle$  of each oscillator is such that  $\omega_{m}+\langle n \rangle \beta  \approx \omega_{m} $, the assumption of  a single resonant frequency for each oscillator  is still a reasonable approximation. Keeping this is mind in the discussion  to follow, we shall restrict ourselves  to low-excitation  subspaces of each oscillator.
 
We are interested in the indirect interaction between the two nonlinear oscillators 
mediated by a linear oscillator with quantized energy levels equispaced by $\hbar \omega$. The indirect coupling is advantageous because it allows for accurate control of the interaction strength by manipulating the mediating oscillator, and consequently  gives a handle on the  quantum dynamics of the two nonlinear oscillators.  The importance of the indirect interactions can be further appreciated in the dissipative regime.  There, if the dissipation rate of the mediating oscillator is much faster than  the thermal relaxation rates of the individual oscillators, then steady state entangled states of the oscillators can be achieved. The linear oscillator is here assumed to be addressable by electromagnetic radiation created  by excess charge or by  nano-magnets at the tip of the oscillators, which produces an oscillating electromagnetic field \cite{ptre}. Making the rotating wave and dipole approximations, the unitary evolution then corresponds to the Hamiltonian
\begin{eqnarray}\label{spcham}
{\widetilde{\hat{H}}_{1}}/{\hbar}& = & \omega_{m}( \hat{a}^{\dagger}\hat{a}+ \hat{b}^{\dagger}\hat{b})+\omega \hat{c}^{\dagger}\hat{c}
+\beta(\hat{n}_{a}^{2}+\hat{n}_{a}+\hat{n}_{b}^{2}+\hat{n}_{b}) \nonumber\\
&&+\kappa(\hat{a}^\dagger \hat{c}+\hat{b}^\dagger \hat{c})+h.c.,
\end{eqnarray}
where $\hat{c}^{\dagger},\hat{c}$ are the creation and annihilation operators for the single quantized mode of the linear oscillator, which couples symmetrically --- with  coupling strength $\kappa$ ---  to each of the nonlinear oscillators. The Hamiltonian \eref{spcham} may, for instance, describe the coherent interaction of two anharmonic oscillators with an ultracold Bose-Einstein condensate (BEC)~\cite{josh}. In which case, in the limit of low atomic excitations, the creation and annihilation operators $\hat{c}^{\dagger}$ and $\hat{c}$ will be analogous to the collective atomic raising and lowering  operators  $\hat{J}^{+}, \hat{J}^{-}$ \cite{dicke,josh}.  As shown in \cite{josh}, the indirect coupling strength $\kappa$ between the two nonlinear oscillators can be made to exceed the direct coupling $\kappa_{\rm direct}$ between them. Moreover, the nonlinearity strength $\beta$ can also be made stronger than the direct coupling strength $\kappa_{\rm direct}$ such that $\kappa_{\rm direct}~<~\beta<~\kappa$. For instance by  following the treatment in \cite{josh} and   treating the two nanocantilevers as anharmonic oscillators  with a zero-point oscillation amplitude of $50~{\rm pm}$ and  with a ferromagnet with $10^{6}$ atoms on each cantilever tip.  If the BEC cloud is treated  as a linear oscillator trapped at a distance $d=1~{\mu}{\rm m}$ above the nanocantilevers and contains $N=10^{4}$ atoms in the trap center then $ \kappa/\kappa_{\rm direct}=8$ and $\beta/ \kappa_{\rm direct}=5$ (see Section 4).

  A general solution of \eref{spcham} may be found, but since the Hamiltonian \eref{spcham} conserves the total number of excitations a significant simplification occurs. If these oscillators can be cooled near to their ground states we can restrict   ourselves to subspaces with few excitations. Furthermore, in order to simplify the analytical and numerical treatment we will truncate the Hilbert space of the middle linear oscillator  to its two lowest excitation subspaces.  
Previously we have found that this assumption results in a rescaling of the Rabi oscillations without altering the qualitative picture \cite{josh}. 

 In the  one-excitation subspace, the relevant basis states for 
 the unitary dynamics governed by  the Hamiltonian \eref{spcham} are  $ |1\rangle_{a}|0 \rangle_{b} |0 \rangle_{c}, |0\rangle_{a}|1 \rangle_{b} |0 \rangle_{c}$ and $|0\rangle_{a}|0 \rangle_{b} |1 \rangle_{c}$. 
 Here $|1\rangle_{a}|0 \rangle_{b} |0 \rangle_{c}$ denotes  a state where one of the anharmonic oscillators is in 
 its first excited state while the other nonlinear oscillator and the linear oscillator  are in their ground states, and analogously for the other combinations. If we assume that the energy splitting of the  linear oscillator  can be  brought in resonance with the oscillation frequency of the anharmonic oscillator, then in the interaction picture the Hamiltonian \eref{spcham} takes the form
 \begin{eqnarray}\label{spcham1}
\hat{H}_{int}&=&\hbar\beta(\hat{n}_{a}^{2}+\hat{n}_{a}+\hat{n}_{b}^{2}+\hat{n}_{b}) 
\nonumber\\&&
+\hbar\kappa(\hat{a}^\dagger \hat{c}+\hat{b}^\dagger \hat{c})+h.c..
\end{eqnarray}
General initial states of the nonlinear-linear coupled oscillator  system in the subspace of one  
excitation can be written as
 \begin{eqnarray}
  |\Psi_1(t)\rangle&=&\sum_{j=0}^1C_{j,1-j,0}(t)|j\rangle_{a}|1-j \rangle_{b} |0 \rangle_{c}\nonumber\\&&
  +C_{0,0,1}(t)|0\rangle_{a}|0 \rangle_{b} |1 \rangle_{c}{\label 4}
 \end{eqnarray}
In the one-excitation subspace, with the initial condition  $|\Psi(0)\rangle= |1\rangle_{a}|0 \rangle_{b} |0 \rangle_{c}$, the time-evolved wave function due to the coherent interactions between the two anharmonic oscillators and the effective
two-level system becomes
\begin{equation}
|\Psi_{1}(t)\rangle = \alpha_{1}(t)|1\rangle_{a}|0 \rangle_{b} |0 \rangle_{c}
+\alpha_{2}(t)|0\rangle_{a}|1 \rangle_{b} |0 \rangle_{c}\nonumber
+\alpha_{3}(t) |0\rangle_{a}|0 \rangle_{b}|1 \rangle_{c},
\end{equation}
where,
 \begin{eqnarray}
\alpha_{1}(t)&=& \bigg(\frac{1}{2}+\frac{e^{-i \beta t/2}}{2}\cos(K_{1}t/2) -i\beta \frac{e^{-i \beta t/2}}{2 K_{1}} 
\sin(K_{1}t/2) \bigg)\\
\alpha_{2}(t)&=&\bigg(-\frac{1}{2}+\frac{e^{-i \beta t/2}}{2}\cos(K_{1}t/2) -i\beta \frac{e^{-i \beta t/2}}{2 K_{1}}\sin(K_{1}t/2) \bigg)\\
 \alpha_{3}(t)&=&-2 i \frac{\kappa}{K_{1}} e^{-i \beta t /2} \sin(K_{1} t/2) |0\rangle_{a}|0 \rangle_{b}|1 \rangle_{c},
\end{eqnarray}
with $K_{1}=  \sqrt{\beta^{2}+8 \kappa ^{2}}$. In the limit  $\beta \rightarrow  0$ we obtain
\begin{eqnarray}
\fl |\Psi_{1}(t)\rangle = \frac{(1+\cos\sqrt{2}\kappa t)}{2} |1\rangle_{a}|0 \rangle_{b} |0 \rangle_{c}+\frac{(-1+\cos(\sqrt{2}\kappa t))}{2}
|0\rangle_{a}|1 \rangle_{b} |0 \rangle_{c}\nonumber\\
-\frac{i}{\sqrt{2}}(\sin(\sqrt{2}\kappa t)) |0\rangle_{a} |0 \rangle_{b} |1 \rangle_{c},
\end{eqnarray}
which coincides with
the wavefunction that describes  
the dynamics of two linear oscillators interacting with an effective two-level system as discussed in \cite{josh}.  It should be noted that the effect of the nonlinearity cannot be fully appreciated in a one-excitation subspace.  In this case the effect of the nonlinearity can be mimicked by making the two oscillators non-resonant 
with the mediating two level system.  Hence to better understand the effect of the intrinsic nonlinearity on the quantum dynamics of each oscillator,  we have to study also the dynamics of the system in the two- and three-excitations subspace.

With the result of the unitary evolution for all three excitation subspaces in hand, we can now attempt to characterize the entanglement between the nano-cantilevers, and by doing so try to 
understand the influence of the inherent nonlinearities. The time-dependent state of the two anharmonic oscillators is a mixed state found by tracing over the  degrees of freedom of the linear oscillator. To quantify the entanglement in a bipartite system in an overall  mixed state, we use the Peres criterion \cite{pere}. We compute the negativity for the reduced density matrix, defined as
\begin{equation}
\label{negativity}
\mathcal N = \frac{1}{2}\left(\sum_{i}^{n}|\lambda_{i}|-1\right),
\end{equation}
where $\sum_{i}^{n}|\lambda_{i}|$ is the sum of the absolute values of all the eigenvalues of the partially transposed reduced density matrix \cite{gvid} of size $n$. A non zero value of negativity $\mathcal N$ ensures inseparability of a bipartite system.

\begin{figure}[!]
\centering
\subfigure[]{
\includegraphics[width=0.41\textwidth]{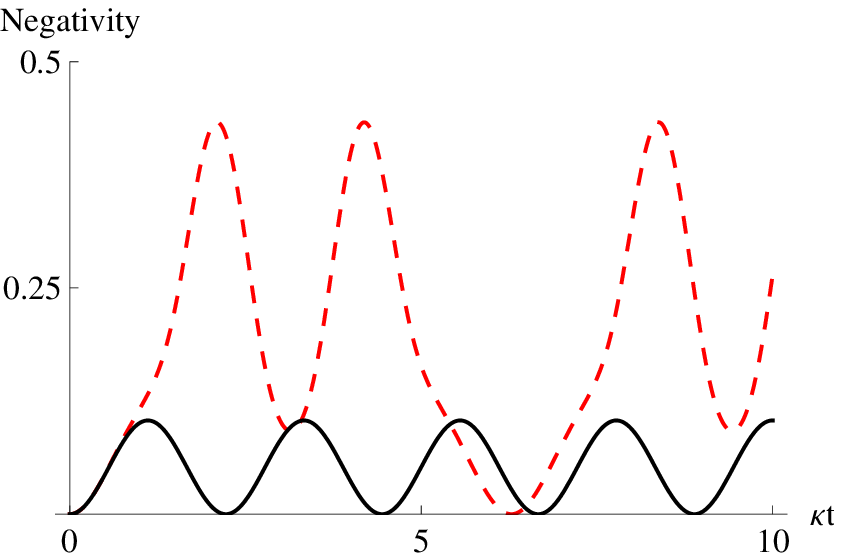}
}
\subfigure[]{
\includegraphics[width=0.41\textwidth]{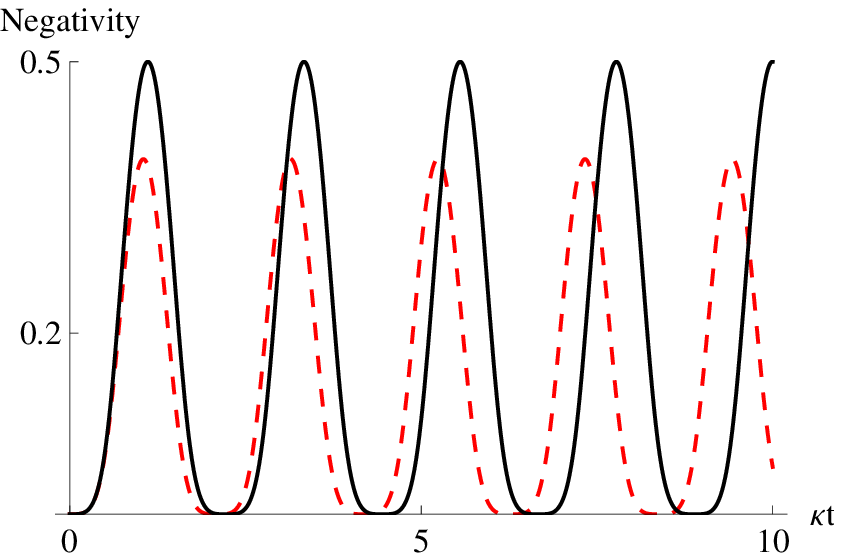}
}
\caption{\label{fig1}(Color online) Degree of entanglement, as measured by the negativity for $\beta/ \kappa=0$ (solid) and $\beta/ \kappa=0.5$ (dashed). The initial states are
 (a) $C_{1,0,0}(0)=1$ (b) $C_{0,0,1}(0)=1$.}
\end{figure}

As a result of the coherent exchange of excitation(s) between the two anharmonic oscillators mediated indirectly via the 
linear oscillator, the two nonlinear oscillators become entangled.  As shown in figure ~\ref{fig1}  the system of two nonlinear oscillators  exhibit time-varying entanglement, and at certain instants the entanglement is found to be maximal or nearly maximal in both excitation subspaces.  
\begin{figure}[!]
\centering
\subfigure[]{
\includegraphics[width=0.41\textwidth]{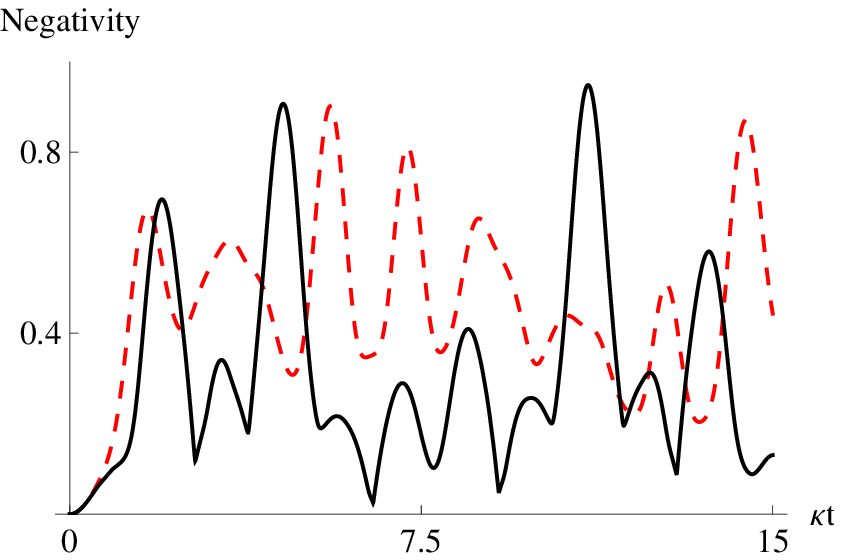}
}
\subfigure[]{
\includegraphics[width=0.41\textwidth]{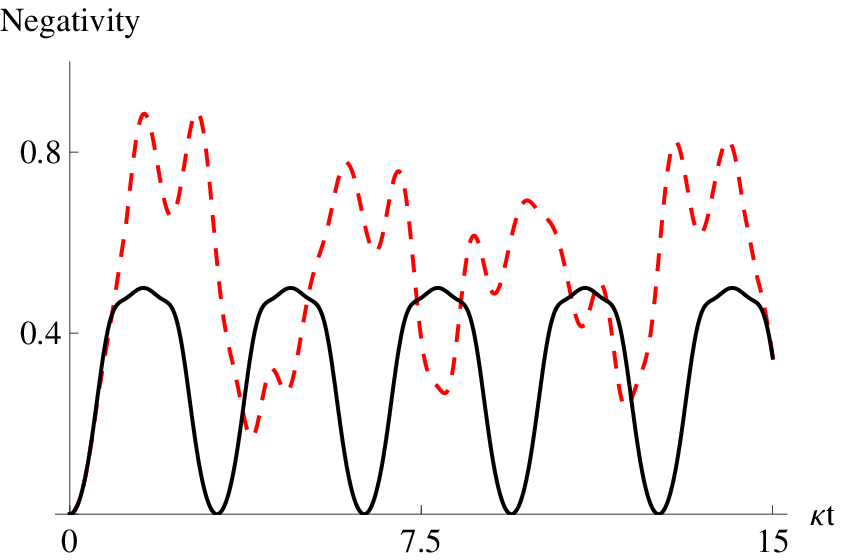}
}
  \caption{\label{fig2}(Color online)  Degree of entanglement, as measured by the negativity for $\beta/ \kappa=0$ (solid) and $\beta/ \kappa=0.5$ (dashed). The initial states are 
 (a) $C_{2,0,0}(0)=1$ (b) $C_{1,1,0}(0)=1$.} 
\end{figure} 
\begin{figure}[!]
\centering
\subfigure[]{
\includegraphics[width=0.41\textwidth]{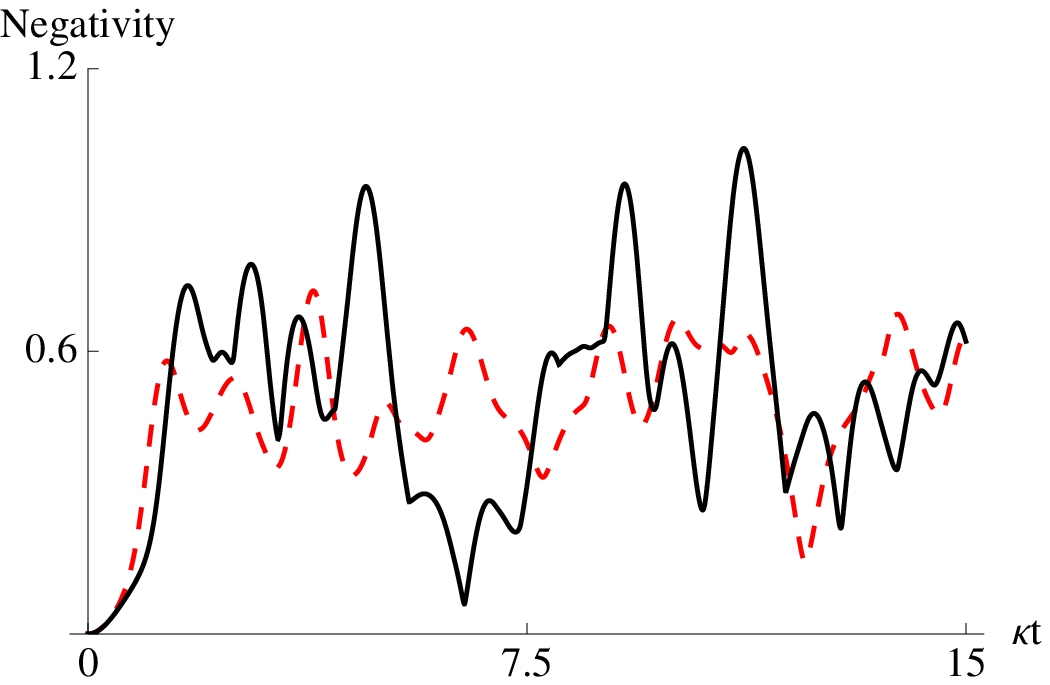}
}
\subfigure[]{
\includegraphics[width=0.41\textwidth]{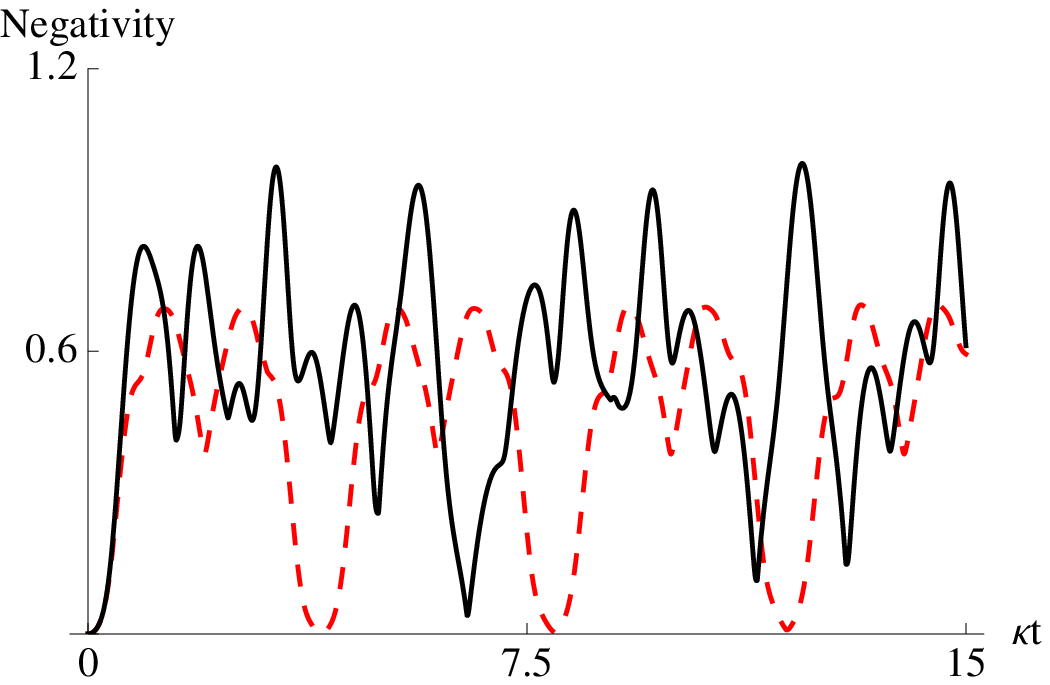}
}
  \caption{\label{fig2newold} (Color online)  Degree of entanglement, as measured by the negativity for $\beta/ \kappa=0$ (solid) and $\beta/ \kappa=0.5$ (dashed). The initial states are 
 (a) $C_{3,0,0}(0)=1$ (b) $C_{1,1,1}(0)=1$.} 
\end{figure} 
For the sake of comparison we also plot the negativity for two indirectly coupled linear oscillators.  As can be seen from figure ~\ref{fig1}, for an initial state given by $|1 \rangle_{a} |0 \rangle_{b} |0 \rangle_{c}$, the nonlinear oscillators exhibit stronger entanglement as compared to their linear counterparts. On the other hand we find that for a symmetric initial state $|0 \rangle_{a} |0 \rangle_{b} |1 \rangle_{c}$, a stronger nonlinearity strength $\beta$ leads to a more weakly entangled state of the two oscillators. 

\begin{figure}[!]
\centering
\subfigure[]
{
\includegraphics[width=0.41\textwidth]{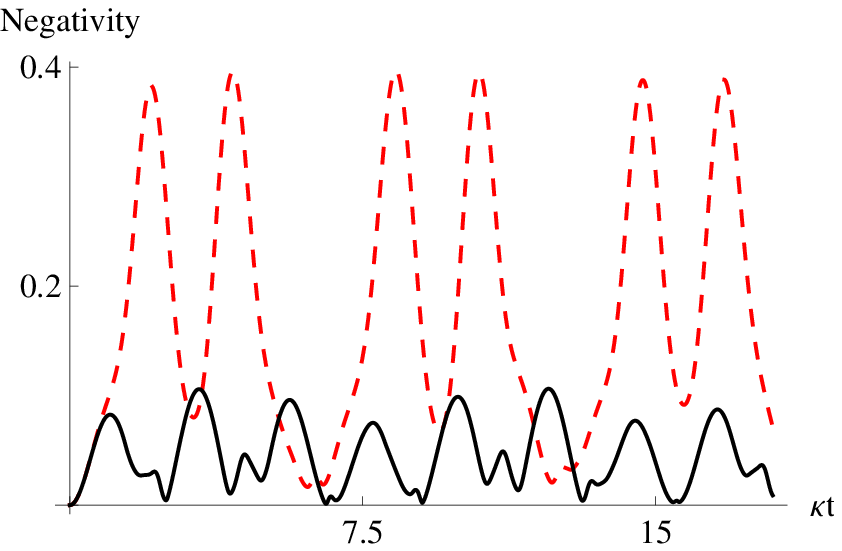}
}
{
\includegraphics[width=0.41\textwidth]{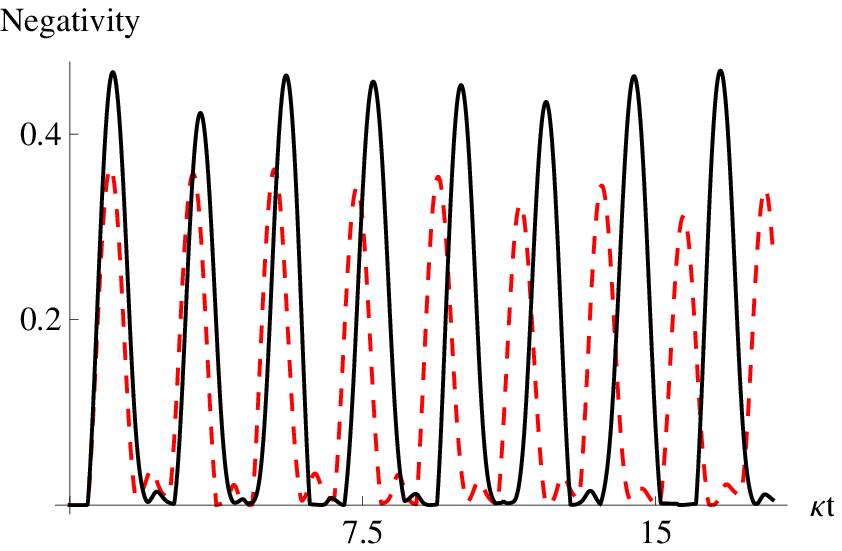}
}
  \caption{\label{figmix} (Color online)  Degree of entanglement as measured by the negativity for  $\beta/ \kappa=0$ (solid) and $\beta/ \kappa=0.5$ (dashed). The initial states are : in 
 (a) a thermal mixture of initial asymmetric states of the three lowest lying excitation subspaces, in (b) a thermal mixture of initial symmetric states of the  three lowest lying excitation subspaces 
 with average thermal occupancy 0.1.} 
\end{figure}

The time evolution of the entanglement for the indirectly coupled nonlinear oscillators in subspaces of higher excitations are shown in figure ~\ref{fig2} and figure ~\ref{fig2newold}.  As can be seen  in subspaces of two and three excitations the effect of a nonlinearity is clearly imprinted on the entangled state of the two oscillators.  All these results indicate that the effect of nonlinearity is much more pronounced for certain initial states. The dynamics becoming more complex in subspaces with higher excitations.  As mentioned before, the particular form of nonlinearity that we are interested in is clearly manifested in subspaces with higher excitations. Adding more excitations to the oscillators will correspond to approaching the semiclassical limit. A more realistic scenario is when the oscillators start in a mixed state. As an illustration of this case, we plot the logarithmic negativity for an initial mixed state of two indirectly coupled nonlinear  oscillators in figure ~\ref{figmix}. As expected, the degree of entanglement is reduced considerably as compared to the case of   initially pure states. Furthermore, crucially depending on the initial state,  a non zero value of $\beta$ may or may not enhance quantum entanglement between the oscillators. 

\subsection{Dissipation of the oscillator}
\noindent
Every physical system is susceptible  to dissipation. A more realistic  approach will therefore take decoherence induced by an environment into account.  Dissipation can either occur through the thermalization of the two anharmonic oscillators or through decay in the effective two-level system. 
As a first approximation we assume that the two nonlinear  oscillators and the linear oscillator are coupled to 
 independent  zero-temperature heat baths with coupling rates  $\gamma_{a,b}$ and $\gamma_{c}$,  respectively, the effect of dissipation on their evolution --- 
under the Born-Markov approximation ---
is well described by a  Lindblad-type master equation of the form 
\begin{equation}\label{13}
\frac{\partial}{\partial t} \hat{\rho} =\frac{-i}{\hbar}[\hat{H}_{int},\hat {\rho}]+\mathcal L_{a}\hat{\rho} +\mathcal L_{b}\hat{\rho}+\mathcal L_{c}\hat{\rho} .
\end{equation}
Here  $\hat{\rho}$ is  the  density matrix of the system, and 
\begin{eqnarray}\label{14}
 \mathcal L_{a}\hat{\rho} &\equiv &\frac{\gamma_{a}}{2} (2\hat{a}\hat{\rho} \hat{a}^{\dagger}-\hat{a}^{\dagger}\hat{a}\hat{\rho}-\hat{\rho} \hat{a}^{\dagger}\hat{a}) \\
 \mathcal L_{b}\hat{\rho} &\equiv &\frac{\gamma_{b}}{2} (2\hat{b}\hat{\rho} \hat{b}^{\dagger}-\hat{b}^{\dagger}\hat{b}\hat{\rho}-\hat{\rho} \hat{b}^{\dagger}\hat{b})\\
 \mathcal L_{c}\hat{\rho}&\equiv &\frac{\gamma_{c}}{2} (2\hat{c}\hat{\rho} \hat{c}^{\dagger}-\hat{c}^{\dagger}\hat{c}\hat{\rho}-\hat{\rho} \hat{c}^{\dagger}\hat{c})
\end{eqnarray}
are  the Lindblad operators representing the coupling of the two anharmonic oscillators and the linear oscillator to their independent zero-temperature heat baths.

\begin{figure}[!]
\centering
\subfigure[]{
\includegraphics[width=0.41\textwidth]{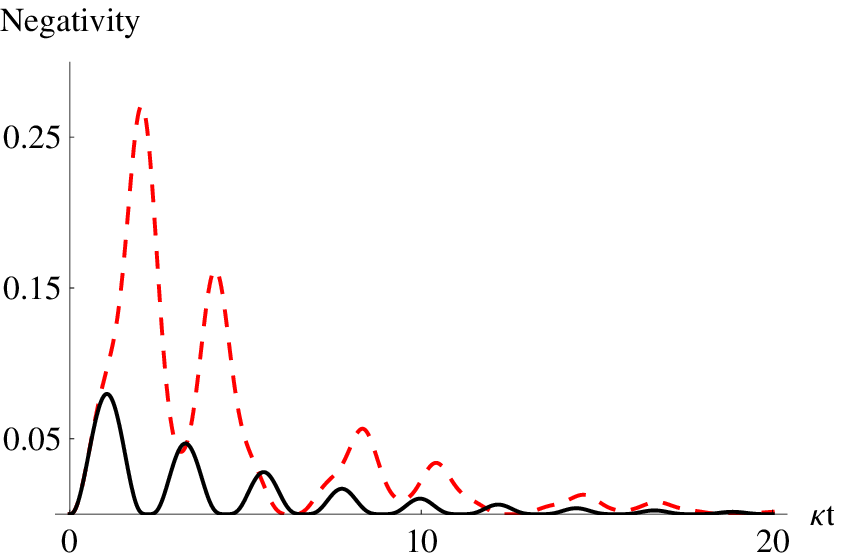}
}
\subfigure[]{
\includegraphics[width=0.41\textwidth]{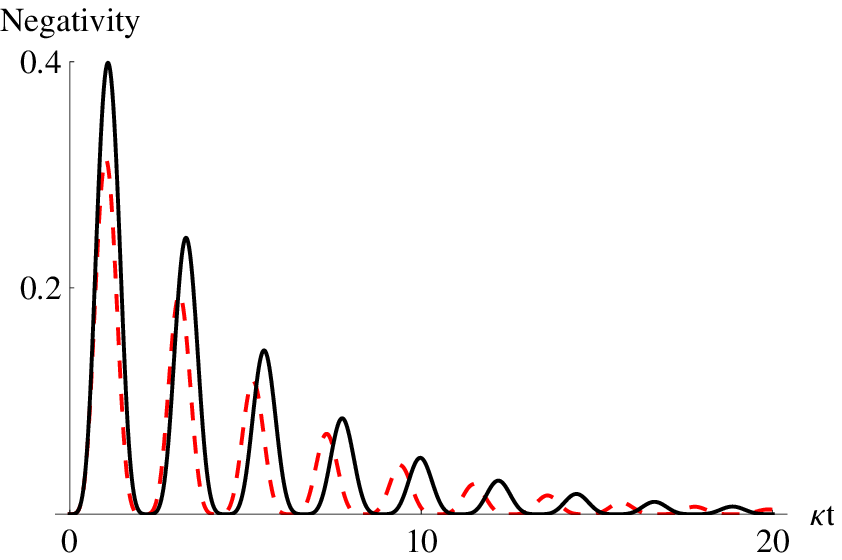}
}
       \caption{\label{fig3}(Color online) Degree of entanglement, as measured by the negativity for $\beta/ \kappa=0$ (solid) and $\beta/ \kappa=0.5$ (dashed) and  $\gamma_{a,b}/ \kappa =\gamma_{c}/ \kappa=0.1$. The initial states are (a) $C_{1,0,0}(0)=1$ and (b) $C_{0,0,1}(0)=1$.}
\end{figure}

A typical numerical solution of \eref{13} in the one-excitation subspace is shown in figure ~\ref{fig3}. As can be seen from the figure, the effect of the nonlinearities is clearly imprinted on the entangled state of the  two oscillators even when they undergo dissipation. As for the case of unitary evolution, the effect of an inherent nonlinearity of the oscillator is much more pronounced for certain initial states.

 The intrinsic nonlinearity of the two oscillators has another dramatic effect  in the sense that it determines  the dissipative dynamics of the coupled oscillators in higher excitation subspaces.  
To see this we solve \eref{13} in the two-excitations subspace. This time we only allow the mediating linear oscillator to undergo dissipation on the time scale of interest. An equivalent problem has been studied in \cite{josh}, where under similar conditions long-lived entangled states of two linear oscillators were achieved.

If $\beta=0$ then $\hat{a}-\hat{b}$ is a constant of motion of the Hamiltonian \eref{spcham1}. Exploiting this fact one can obtain steady states of the two linear oscillators which are entangled \cite{josh}.  
On the other hand, for nonlinear oscillators this does not hold true. Here, depending on the initial state, one may or may not see a steady-state entangled state of the two nonlinear oscillators develop.
 
\begin{figure}[!]
\centering
\subfigure[]{
\includegraphics[width=0.41\textwidth]{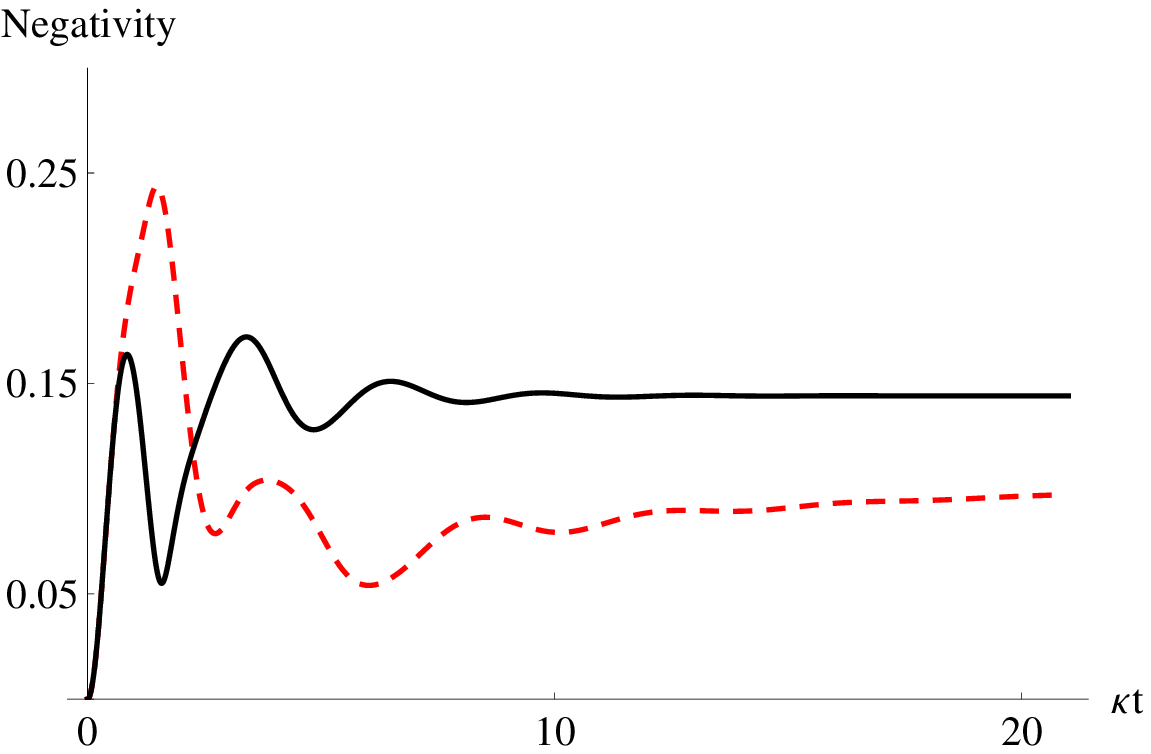}
}
\subfigure[]{
\includegraphics[width=0.41\textwidth]{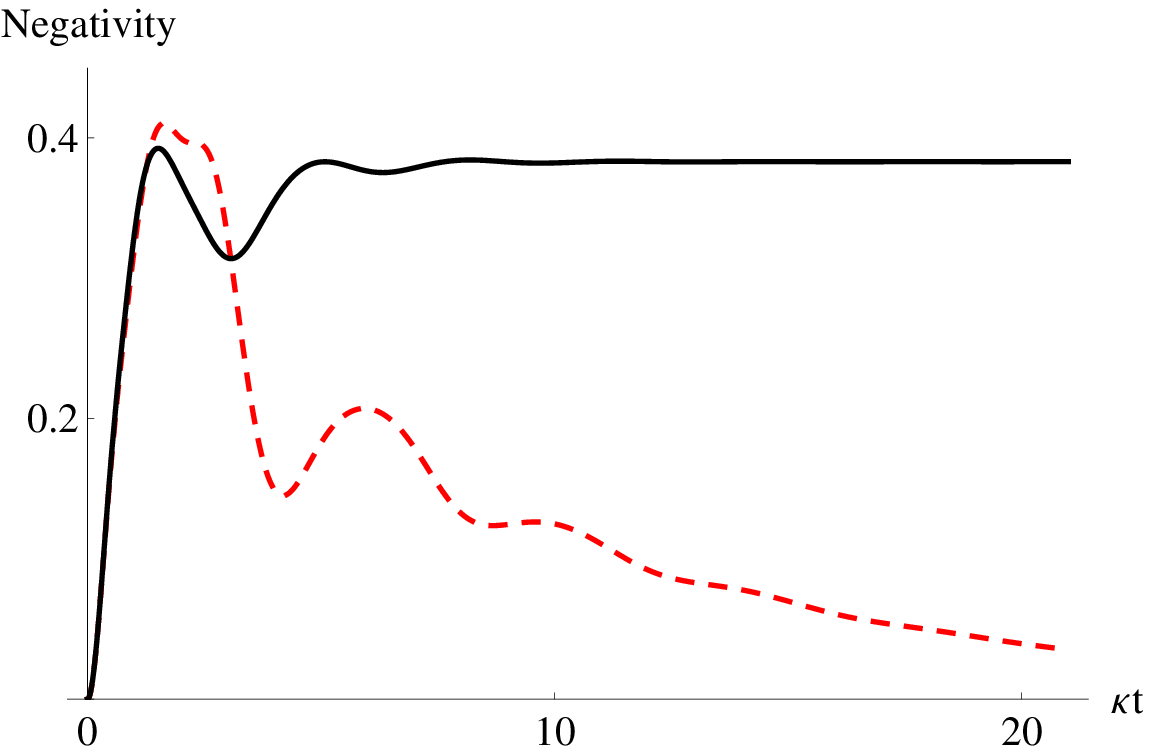}
}
 \caption{\label{fig4}(Color online) (a) Degree of entanglement, as measured by the negativity for $\beta/ \kappa=0$ (solid) and $\beta/ \kappa=0.5$ (dashed),  $\gamma_{c}/ \kappa =2$ and $\gamma_{a,b}/ \kappa =0$.  The initial states are 
(a) $C_{2,0,0}(0)=1$ and (b) $C_{1,1,0}(0)=1$ .}
\end{figure}

To prove this statement a numerical solution of \eref{13} is shown in figure ~\ref{fig4}.  As can be seen,  for  an initial asymmetric state the steady state is an entangled state of the two nonlinear oscillators while for an initial symmetric state the steady state is separable.  These observations can thus also be used as an indirect signature of the state of the nonlinear oscillator.

As is clear from this section the degree of inseparability of the two indirectly coupled oscillators  has a non trivial dependence on both the  nonlinearity parameter $\beta$ and the initial state. Depending on the initial state a non zero value of nonlinearity strength $\beta$ can enhance or suppress the degree of entanglement of  indirectly coupled oscillators. This interesting behaviour holds true both when the system undergoes unitary evolution and  in the dissipative regime.

 \section{Interaction with a quantized cavity mode}
 \label{sec:cavity}
\noindent
In the previous section we saw how an intrinsic nonlinearity can strongly affect the entanglement between the oscillators.
Here we will discuss a second physical scenario where nonlinearities are playing a key role in the quantum dynamics. The first setup we have in mind is an anharmonic oscillator  coupled to a mode of a Fabry-P\'erot cavity with one fixed and one movable mirror. 

In \cite{sman,sbos} the problem of a cavity with a movable mirror has been discussed in great detail. In these studies the movable mirror was treated as a simple harmonic oscillator.   This work was purely analytical,  and  
showed that the coherent interaction of a movable mirror  with the cavity mode 
 generates various non-classical states of the cavity mode and the mirror.
Here we are interested in probing the quantum features of an anharmonic oscillator. We shall model the movable mirror as a nonlinear oscillator with a nonlinearity proportional to $x^{4}$, where $x$ is the displacement of the mirror from its equilibrium position.  In what follows we study the coherent interaction between a single quantized cavity mode and  a nonlinear mirror 
coupled by the radiation pressure \cite{afpa}.  We derive a closed analytical expression for the time-evolved state of the cavity field and the movable anharmonic mirror, which is valid in the limit of weak nonlinearity and low radiation pressure coupling.

If we assume that leakage of photons through the cavity can be neglected, then the main source of decoherence is the coupling of the mirror to its surroundings, which to some extent can also be avoided \cite{ssinnew}. In what follows we therefore neglect dissipation and only consider unitary evolution of the system of  of the coupled cavity- and nonlinear-mirror system. 

Consider a single quantized cavity  mode with  creation and annihilation   operators $ \hat{k}^{\dagger}$ and $\hat{k}$, and resonance frequency $\omega_{k}= 2\pi c  /L$, where $L$ is the length of the cavity. We assume that the movable mirror has been cooled near to its ground state and thus is operating in its quantum regime. Under the action of cavity photon induced radiation pressure, the movable mirror will oscillate about its equilibrium position. If we
assume that the mirror   moves a distance $x$ along the cavity axis such that the displacement is much smaller than the wavelength of the cavity mode in one cavity round-trip time, then the scattering of photons to other cavity modes can be safely neglected \cite{afpa}. 
The length of the cavity then becomes $L+x$ so that the resonance frequency of the cavity  is of the form  $\omega_{k}^{\prime}=  2\pi c  /(L+x)$. The Hamiltonian of the cavity can then be rewritten as
\begin{equation}
\hat{H}_{cav}=\hbar \omega_{k}^\prime\hat{k}^{\dagger}\hat{k}
=2\pi\hbar\,\frac{ c } {L+x}\,\hat{k}^{\dagger}\hat{k},
\end{equation}
which, in a quantum description of the mirror motion, becomes
\begin{equation}
\hat{H}_{cav}=\hbar \omega_{k} \hat{k}^{\dagger}\hat{k}-\hbar g_{k} \hat{k}^{\dagger}\hat{k} (\hat{a}^{\dagger}+\hat{a}),
\end{equation}
where it is assumed that $x/L\ll1$, and $g_{k}=(\omega_{k}/L)\sqrt{\hbar/2m\omega_{m}}$ is the radiation pressure coupling constant  between the nonlinear mirror and the cavity field. Thus the unitary dynamics of the above physical system  is governed by the
Hamiltonian
   \begin{equation}\label{hamcav1}
{\hat{H}_{2}}/{\hbar}=\omega_{k}\hat{k}^{\dagger}\hat{k}+(\omega_{m}+\beta) \hat{a}^{\dagger}\hat{a}+\beta(\hat{a}^{\dagger}\hat{a})^{2}-g_{k}\hat{k}^{\dagger}\hat{k}(\hat{a}^{\dagger}+\hat{a}),
\end{equation}
where the nonlinear mirror has been approximated by a quartic anharmonicity as in \eref{hamnonline11}. 
The Hamiltonian in  \eref{hamcav1} can be rewritten using the transformation
\begin{equation}
\label{hamcav2}
\hat{H}_{\rm trans}=e^{\hat{S}}\hat{H}_{2}e^{-\hat{S}},
\end{equation}
where the unitary operator $\hat S$ is given by
\begin{equation}\label{opert}
\hat{S}=-\frac{g_{k}}{\omega_{m}+\beta}\hat{k}^{\dagger}\hat{k}(\hat{a}^{\dagger}-\hat{a}).
\end{equation}
Consequently the operators $\hat{a}$ and $\hat{k}$ transform as
\begin{eqnarray}\label{hamcav4}
\hat{a} &\rightarrow&  \hat{a}+\frac{g_{k}}{\omega_{m}+\beta}\hat{k}^{\dagger}\hat{k}, \\
\hat{k} &\rightarrow& \hat{k} \exp\left[{\frac{g_{k}}{\omega_{m}+\beta}(\hat{a}^{\dagger}-\hat{a})}\right].
\end{eqnarray}

Neglecting the counter-rotating terms, the transformed Hamiltonian in  \eref{hamcav2} becomes
\begin{eqnarray}\label{hamcav5}
\frac{\hat{H}_{\rm trans}}{\hbar}&=&\omega_{k} \hat{k}^{\dagger} \hat{k}+(\omega_{m}+\beta)\hat{a}^{\dagger}\hat{a}-\frac{g_{k}^{2} \omega_{m}}{(\omega_{m}+\beta)^{2}} (\hat{k}^{\dagger}\hat{k})^{2}\nonumber\\ &&
+\beta(\hat{a}^{\dagger}\hat{a})^{2}+\frac{4g_{k}^{2}\beta}{(\omega_{m}+\beta)^{2}}(\hat{a}^{\dagger} \hat{a})(\hat{k}^{\dagger} \hat{k})^{2}\\&&
+2\beta{( \frac {g_{k}}{\omega_{m}+\beta})}^{3}(\hat{k}^\dagger \hat{k})^{3}(\hat{a}+\hat{a}^{\dagger})+\frac{g_{k}^{4} \beta (\hat{k}^{\dagger}\hat{k})^{4}}{(\omega_{m}+\beta)^{4}}\nonumber.
\end{eqnarray}
 To further simplify the analysis we  assume that both the nonlinearity and the radiation-pressure coupling are  weak, so that quadratic  and higher orders terms in  $g_{k}/(\omega_{m}+\beta)$ can be neglected. This can be justified  since 
 a cavity of length $L\sim 10^{-3}$~m and a movable mirror with oscillation frequency $\omega_{m}\sim 10^6$~Hz and  zero-point oscillation amplitude
 50 pm gives $g_{k}/\omega_{m} \sim 0.01$.  Thus \eref{hamcav5} reduces to
\begin{equation}\label{hamcav5new}
\frac{\hat{H}_{\rm trans}}{\hbar}=\omega_{k} \hat{k}^{\dagger} \hat{k}+(\omega_{m}+\beta)\hat{a}^{\dagger}\hat{a}
 -\frac{g_{k}^{2} \omega_{m}}{(\omega_{m}+\beta)^{2}}(\hat{k}^{\dagger}\hat{k})^{2} +\beta(\hat{a}^{\dagger}\hat{a})^{2}.
\end{equation}
It should be noted that  $\hat{n}_{k}=\hat{k}^{\dagger}\hat{k}$ and $\hat{n}_{a}=\hat{a}^{\dagger}\hat{a}$ are constants of motion  since $[\hat{H}_{\rm trans},\hat{n}_{k}] = [\hat{H}_{\rm trans},\hat{n}_{a}] =0$.
The transformed unitary time-evolution operator corresponding to 
$\hat H_{\rm trans}$ takes the form
\begin{eqnarray}\label{hamcavunit}
\hat{U}_{\rm trans}(t)&=&\exp[{-i\omega_{k}t \hat{k}^{\dagger} \hat{k}+i\frac{g_{k}^{2} \omega_{m}}{(\omega_{m}+\beta)^{2}}t (\hat{k}^{\dagger}\hat{k})^{2}}]\quad\quad\\
&&\times\exp[{-i(\omega_{m}+\beta)t\hat{a}^{\dagger}\hat{a}-i\beta t(\hat{a}^{\dagger}\hat{a})^{2}}].
\nonumber
\end{eqnarray}
The corresponding untransformed time evolution operator is $\hat{U}(t)=e^{-\hat S}\hat{U}_{\rm trans}(t)e^{\hat S}$.
See the Appendix for technical details regarding the exact form of $\hat U(t)$.

\begin{figure}[!]
\centering
\subfigure[]{
\includegraphics[width=0.41\textwidth]{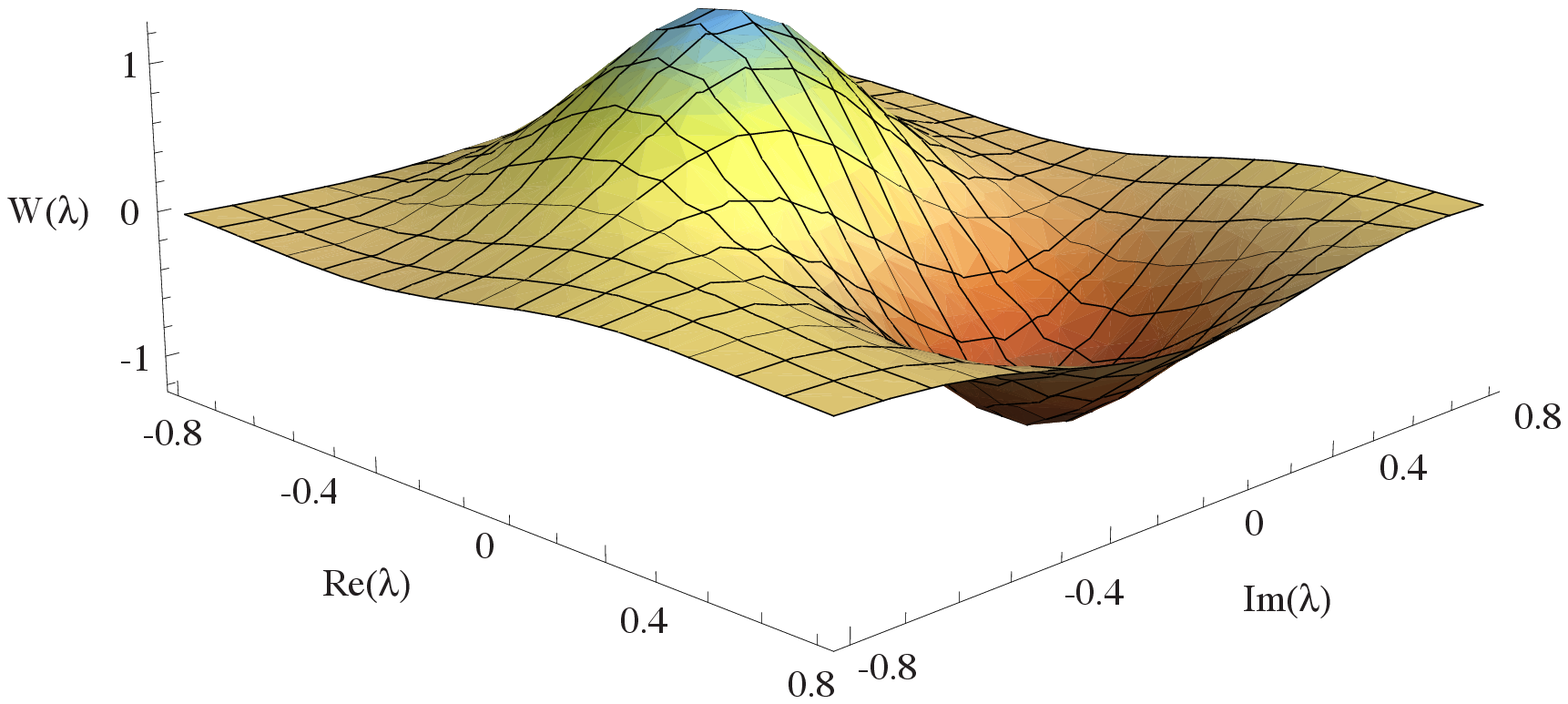}
}
\subfigure[]{
\includegraphics[width=0.41\textwidth]{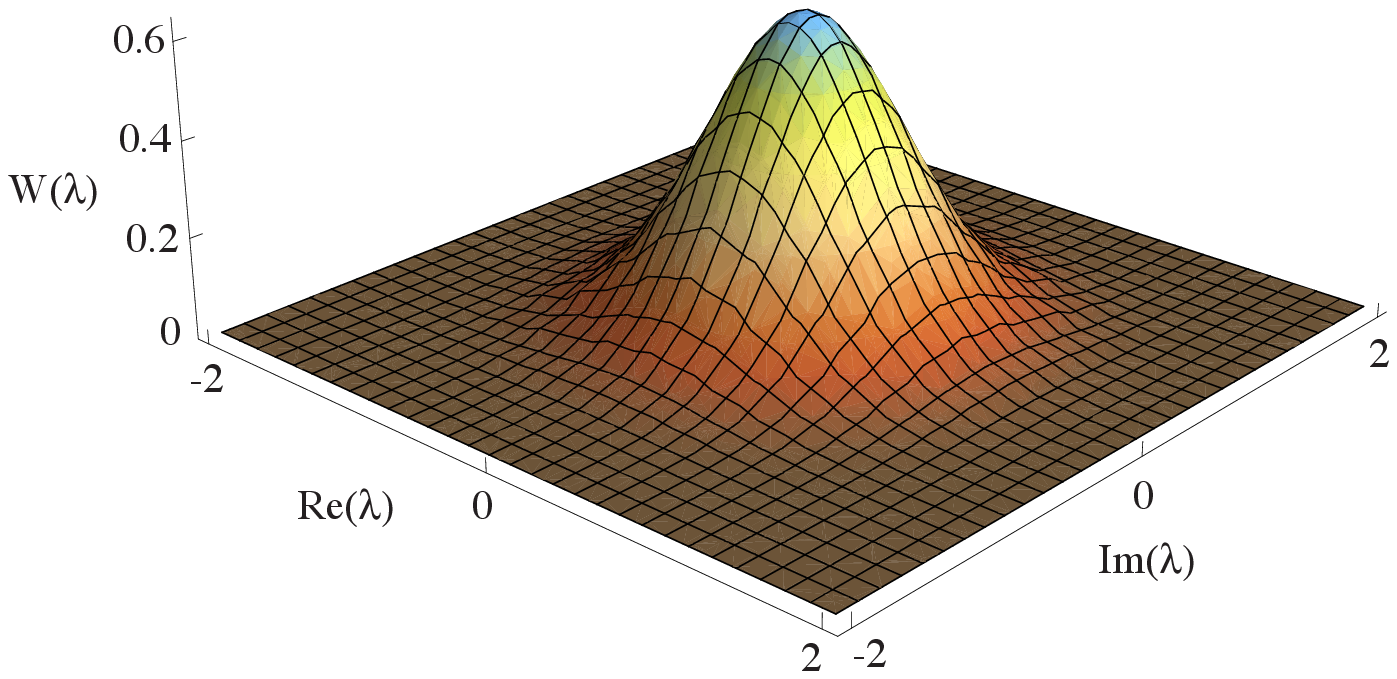}
}
  \caption{\label{wigner}(Color online)  (a) Wigner function of the movable mirror initially prepared in its ground state and interacting with a cavity mode with(a) $\beta/(\omega_{m}+\beta)=10^{-4}$ (b) $\beta/(\omega_{m}+\beta)=0$. Initially $|\alpha|^{2}=1$;  $g_{k}/(\omega_{m}+\beta)=10^{-2}$ and $(\omega_{m}+\beta)t=\pi/$4.} 
\end{figure}

\begin{figure}[!]
\centering
\subfigure[]{
\includegraphics[width=0.41\textwidth]{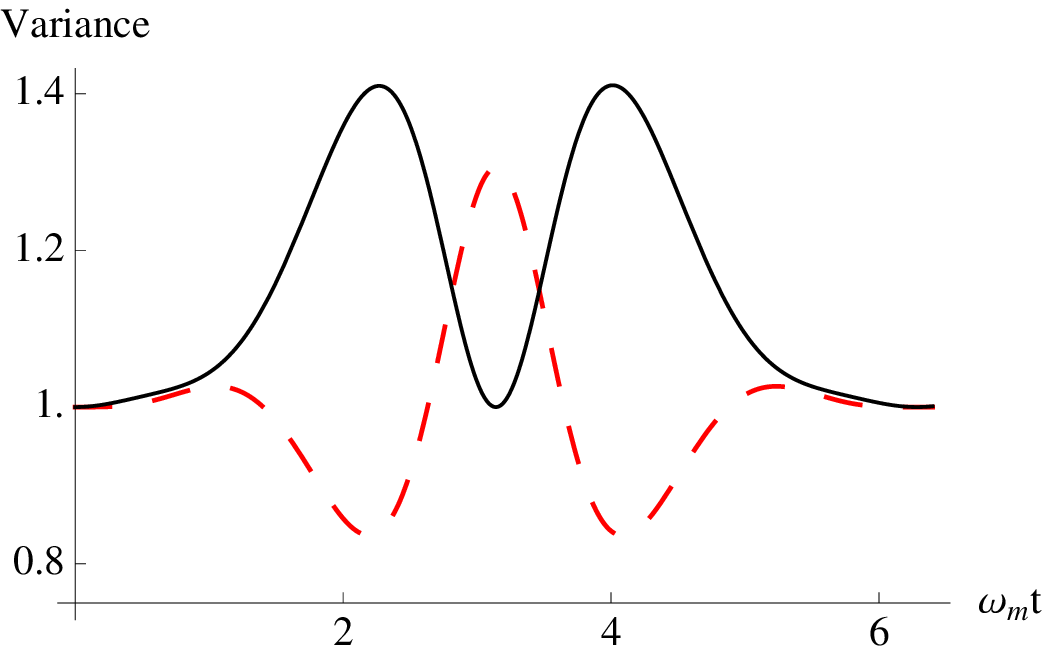}
}
\subfigure[]{
\includegraphics[width=0.41\textwidth]{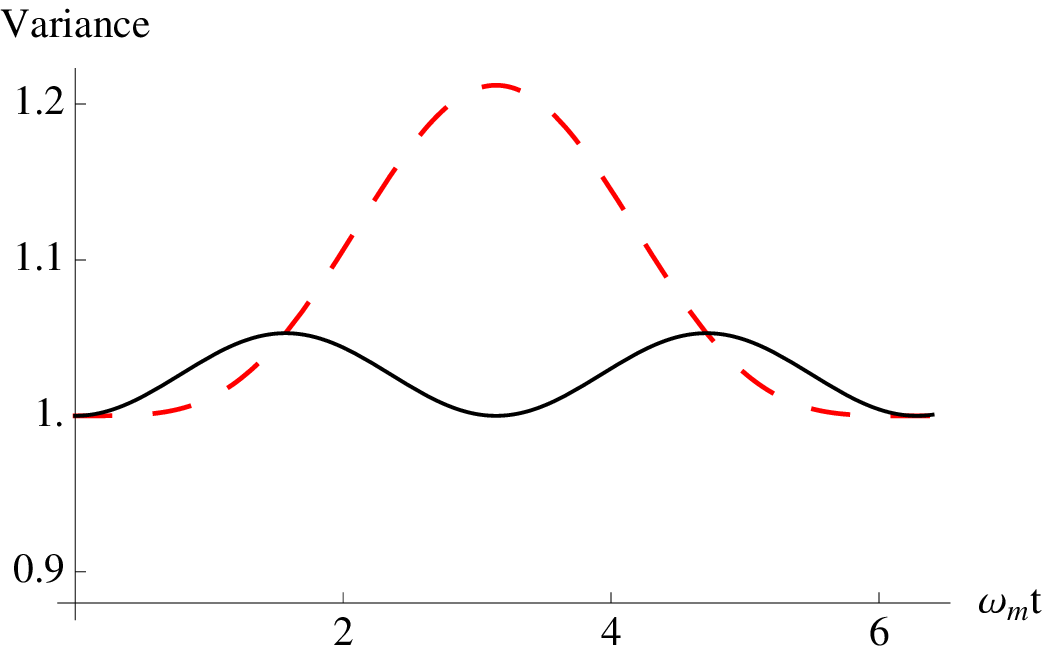}
}
  \caption{\label{mirrorquadnonlnear1}(Color online) Time variation of the variance of the mirror quadratures $\hat{P}$ (solid) and $\hat{Q}$ ( dashed) with the mirror initially prepared in its vacuum state, where $g_{k}/(\omega_{m}+\beta)=0.06$ and  $|\alpha|^2 =5$. (a) $\beta/ (\omega_{m}+\beta) =10^{-4}$ and (b) $\beta/ (\omega_{m}+\beta) =0$.   As a result of coherent interactions with a cavity mode an  anharmonic oscillator exhibits time dependent squeezing beyond the minimum uncertainty limit in one of its quadratures.} 
\end{figure}

Under the assumption of weak nonlinearity and low radiation pressure coupling, $\hat U(t)$ describes the undamped motion of an anharmonic oscillator interacting with a cavity mode. If we assume that both the cavity mode and the oscillator are prepared in coherent states with  amplitudes $\alpha$ and $\eta$ respectively, then the state of the combined system evolves as
\begin{eqnarray}\label{stateevol}
|\Psi(t)\rangle&=&\hat{U}(t)|\Psi(0)\rangle \nonumber\\
&=&\sum_{n=0}^{\infty} e^{-|\alpha|^2/2} \frac{\alpha^n}{\sqrt{n!}} e^{i \left[(\frac{g_{k}}{\omega_{m}+\beta})^2 n^2 (\omega_{m}t-\sin(\omega_{m}+\beta)t\right]}\nonumber\\
&&\times e^{-in\omega_{k}t}  |n \rangle_{c}| \tilde{\eta}(t) \rangle_{a},
\end{eqnarray}
where the state of the mirror is a mixture of Fock states given by
\begin{equation} \label{anhrcohevl}
| \tilde {\eta}(t)\rangle_{a}= \sum_{m=0}^\infty [\tilde{ \eta}(t)]^{m} e^{-|\tilde {\eta}(t)|^{2}/2} e^{-i\beta  t m^2}\frac{1}{\sqrt{m !}}|m\rangle_{a},
\end{equation}
with  
\begin{equation}
\tilde{ \eta}(t)=\eta e^{-i(\omega_{m}+\beta) t}+\frac{g_{k}}{\omega_{m}+\beta}n[1-e^{-i(\omega_{m}+\beta)t}].
\end{equation}
In the limit $\beta  \rightarrow 0$ we retrieve the result obtained in \cite{sman, sbos} where the mirror state reduces to a mixture of coherent states. It is worth noting that even in the weakly nonlinear regime the effect of the nonlinearity is clearly imprinted on the state of the movable mirror which is now in a mixture of Fock states. Also evident from \eref{anhrcohevl} is the inseparable state of the nonlinear oscillator and the cavity mode. As can be seen from \eref{stateevol}, the anharmonic oscillator exhibits periodic entanglement with the cavity mode, except at certain instants where the state of the oscillator is completely separable from the cavity mode. This  happens when $(\omega_{m}+\beta)t =2 q \pi$ for $q \in \mathbb{N}$. The reduced density matrix of the state of the mirror is thus given by,
\begin{eqnarray} \label{rhonewhai}
\hat{\rho}_{\rm{mirror}}(\rm{t})=Tr_{k}(|\Psi(\rm{t})\rangle \langle \Psi(\rm{t})|)=   \sum_{m,q,n=0}^\infty e^{-|\alpha|^2}|\alpha|^{2n} \\
\frac{1}{\sqrt{n!}}(\tilde{ \eta}(t))^{m}   (\tilde{ \eta^{*}}(t))^{q} e^{-|\tilde {\eta}(t)|^{2}} e^{i\beta  t(q^2- m^2)}\frac{1}{\sqrt{m !}}\frac{1}{\sqrt{q !}}|m\rangle_{a}\langle q|.
\end{eqnarray}
 It is now a straightforward exercise to compute the Wigner  function $W(\lambda,\lambda^{*})$ \cite{stev} of the mirror, which we plot in figure~\ref{wigner}. As can be seen from this figure the negativity of the Wigner function $W(\lambda,\lambda^{*})$ clearly identifies the non-classical state of the mirror. It should be contrasted with the case of a linear oscillator interacting with a cavity mode. There the state of the mirror is a mixture of coherent states and always characterized by a positive Wigner function. The evolution of the mirror into a non-classical state is an effect of the combination of an intrinsic nonlinearity of the mirror and the radiation pressure coupling with the cavity mode. This feature should be compared with results obtained in \cite{sbos} where it has been shown that only a conditional measurement on the cavity mode can project the linear mirror into a non-classical state.
One would expect the amplitude and the phase quadratures of the movable mirror to be influenced by the intrinsic nonlinearity in the mirror. In order to quantify this we define two Hermitian operators $\hat{Q}$ and $\hat{P}$, which correspond to the amplitude and phase quadratures of the movable mirror and are given by
\begin{eqnarray}\label{quadrature}
\hat{Q}(t)&=&(\hat{a}^{\dagger}(t)+\hat{a}(t)),\\
\hat{P}(t)&=&i(\hat{a}^{\dagger}(t)-\hat{a}(t)). 
\end{eqnarray}
The coherent interaction of the cavity with the anharmonic oscillator should be reflected in the variance of the quadratures $\hat{P}$ and $\hat{Q}$ defined by
\begin{eqnarray}
\Delta \hat{P}^{2}(t) = \langle \hat{P}^{2}(t) \rangle-\langle \hat{P}(t)\rangle^{2}, \\ 
\Delta \hat{Q}^{2}(t) = \langle \hat{Q}^{2}(t) \rangle-\langle \hat{Q}(t)\rangle^{2}.
\end{eqnarray}
We analytically solve for $\hat{P}^2(t)$ and $\hat{Q}^2(t)$ and plot the variance of the quadratures $\hat{P}$ and $\hat{Q}$ in figure~\ref{mirrorquadnonlnear1}. As can be seen there, 
the coherent interaction between a quantized cavity mode and an anharmonic oscillator induces a time-dependent squeezing in one of the the mirror  quadrature beyond the minimum uncertainty limit.

It  is worth pointing out that the squeezing in the variance of the mirror quadratures beyond the minimum uncertainty limit is the result of a combined effect of the intrinsic nonlinearity and the radiation pressure coupling. This can be understood from the fact that if the mirror is initially prepared in its vacuum state then it is known that a nonlinearity of the form \eref{hamnonline11}  alone cannot induce squeezing in the mirror quadratures \cite{ajij}.  As a result of joint coherent interaction with the cavity mode and intrinsic nonlinearity an initial vacuum state of the nonlinear mirror evolves into a mixture of Fock states which exhibit time dependent  squeezing beyond the minimum uncertainty limit.

\section{Origin of the nonlinearities}
\label{sec:out}
\noindent
The harmonic oscillator is often the result of an approximation of a more complicated potential landscape. Nonlinear force terms are often naturally present in many physical systems, but they are of higher order, hence small.
 Here we shall outline a physical scheme for inducing nonlinearity of a mechanical oscillator in the form of a nano-cantilever \cite{canti}, where the nonlinear quartic term appears as a lowest order approximation.
In order to obtain a nonlinear oscillator, we propose to use a hybrid system which relies on the electromagnetic coupling between nano-magnets located at the tip of the cantilever and external magnetic fields.
Consider a setup consisting of two identical circular magnetic coils of radii $R$ placed a distance of $R/2$ apart, with their common axis along the $x$ direction. This Helmholtz coil configuration is known to produce a very uniform magnetic field near the centre.  A nanocantilever  cooled near to its vibrational ground state and fabricated with a strong ferromagnet of magnetic moment $\vec{\mu}$ attached to its tip is placed at the centre of the Helmholtz coil setup. 
 
The magnetic field experienced by the ferromagnet at the tip of the nanocantilever, placed at the centre of the Helmholtz  coils, is given by
\begin{equation}
B(x)=\frac{\mu_{0}n_{\rm turns}I}{2R}\left\{\left[1+\frac{(\frac{R}{2}-x)^2}{R^2}\right]^{-\frac{3}{2}}\right.
+\left.\left[1+\frac{(\frac{R}{2}+x)^2}{R^2}\right]^{-\frac{3}{2}}\right\},
\label{mag}
\end{equation}
  where $x$ is the displacement of the tip of the oscillator from the centre  and  $I$ is the current in the pair of coils. Simplifying \eref{mag} for $x/R<<1$ we get
\begin{equation}\label{mag1}
B(x)=\frac{8\mu_{0}I}{5R\sqrt{5}}\left[1-\frac{144}{125}\left(\frac{x}{R}\right)^{4}\right].
\end{equation}
Thus the interaction energy of the ferromagnet is given by
\begin{equation}\label{int}
H_{int}= -\mu\cdot B(x) 
\approx \frac{0.8\mu_{0}\mu I} {R} \left({\frac{x}{R}}\right)^{4}.
\end{equation}
 Representing the quantized motion of the oscillator in terms of  creation and  annihilation  operators $\hat{a}^{\dagger}$ and $\hat{a}$, the interaction Hamiltonian takes the form
\begin{equation}\label{int1}
H_{int} = \hbar \beta (\hat{a}+\hat{a}^{\dagger})^4,
\end{equation}
where $\beta=1.28\mu_{0} \mu_{B} N_{\rm mag} I {a_{0}}^4/\hbar R^{5}$, $N_{\rm mag}$ is the number of atoms in the ferromagnet, and $a_{0}$ is the zero point amplitude of the nanomechanical cantilever.

Using the  physical setup described above, a nonlinearity of strength $\beta$ can be induced in the nanomechanical oscillator provided the zero point motion of the cantilever can be made large (see \cite{reviewartcle} for a review of the present state of the art manufacturing of nanomechanical oscillators). For a set of parameters where $R\sim 80$~nm, $I \sim 1$~mA, $N_{\rm mag}\sim10^{6}$, and $a_{0}\sim 50$~pm, one obtains a nonlinearity strength of the order of $\beta \sim 250$~Hz, where we have neglected any finite size effects stemming form the nanomagnet and coils.

\section{Conclusions}  
\label{sec:concl}
\noindent
We have studied the dynamics of anharmonic oscillators with a quartic nonlinearity in two different physical settings. 
We have described, in detail, the quantum evolution of two such anharmonic oscillators interacting indirectly via an effective two-level system.  The two-level system could also, for example, be represented by some chosen
collective excitations of a Bose-Einstein condensate.
We have shown that indirect coherent interactions cause the two anharmonic oscillators to exhibit time-dependent entanglement. Inherent nonlinearities in the nano-mechanical systems are found to strongly influence the entangled state of the two oscillators. Interestingly, the effect of nonlinearity is much more pronounced for certain initial states. The signature of nonlinearity is clearly imprinted on the entangled state of the two anharmonic oscillators even when these oscillators are subject to decoherence. Nonlinearity also plays a crucial role in determining the steady state evolution of the indirectly coupled harmonic oscillators.

The coupling strength between two oscillators can be characterized by the connectivity  \cite{lcho}. Connectivity as defined in \cite{lcho} is the ratio of the coupling strength between the oscillators and the frequency difference between them, and diverges in the limit of identical oscillators.  A high value of the connectivity corresponds to coherent exchange of excitations between the oscillators, which is desirable in order to operate in the strong coupling regime where coherent  interactions supersede all the losses in the system. In the particular physical model studied here, however, we have found that a larger value of the coupling strength does not always guarantee a strongly entangled state of the two oscillators. The strength of these quantum correlations also depends on the nonlinearity parameter and the initial state distribution. In addition, a very large coupling strength also makes the rotating wave approximation questionable.

As a second illustration of the effect of nonlinearity, 
we have studied  the coherent interaction between a single quantized cavity mode and a weakly nonlinear oscillator in the form of a movable mirror. In this case we have been able to find an analytical solution for the unitary evolution of the state of such an anharmonic oscillator interacting with a single quantized cavity mode. In particular, we have shown that non-classical states of the mirror arise as a result of the combination of the radiation pressure coupling and the intrinsic nonlinearity in the mirror. A non-classical state of the mirror can be generated both for initial ground and coherent states. Unlike in previously studied cases, non-classical states of the mirror can be generated without the need of conditional measurements  \cite{sman,sbos}.  In addition we have shown how squeezing appears in the variance of the quadratures beyond the minimum uncertainty state. It should be stressed that for an initial ground state of a single nonlinear mirror no squeezing will be generated. Squeezing only occurs due to the interaction between the nonlinear mirror and the cavity field.
\ack
CJ gratefully acknowledges financial support by the ORS scheme and MJ partial support from the Korean WCU program funded by MEST through the NFR (Grant No. R31-2008-000-10057-0).

\section*{Appendix}
\section{The evolution operator}
\noindent
The unitary operator ${{\hat S}}$ in \eref{opert} which is used to transform the Hamiltonian $\hat H_{2}$  in \eref{hamcav5new} gives the corresponding transformed time evolution operator
\begin{eqnarray}\label{hamcavunit}
\fl \hat{U}_{\rm trans}(t)=\exp\left[{-i\omega_{k}t \hat{k}^{\dagger} \hat{k}+i\frac{g_{k}^{2} \omega_{m}}{\zeta^{2}}t (\hat{k}^{\dagger}\hat{k})^{2}}\right]\nonumber\\
\exp\left[{-i\zeta t\hat{a}^{\dagger}\hat{a}-i\beta t(\hat{a}^{\dagger}\hat{a})^{2}}\right],
\end{eqnarray}
where $\zeta=\omega_{m}+\beta$.
The untransformed operator $\hat U(t)$ then becomes 
\begin{eqnarray}\label{hamcavunit1}
\hat{U}(t)=e^{-\hat S}{{\hat{U}_{\rm trans}}(t)}e^{\hat S}\nonumber\\
=\exp\left[{-i\omega_{k}t \hat{k}^{\dagger} \hat{k}+i\frac{g_{k}^{2} \omega_{m}}{\zeta^{2}}t (\hat{k}^{\dagger}\hat{k})^{2}}\right] \nonumber\\
\exp({-\hat{S} })\exp\left[{-i\zeta t\hat{a}^{\dagger}\hat{a}-i\beta t(\hat{a}^{\dagger}\hat{a})^{2}}\right]\exp({\hat{S}}).
\end{eqnarray}
Using the Baker-Campbell-Hausdorf expansion \cite{stev} together with making the rotating wave approximation, and also neglecting quadratic and higher order terms in  $g_{c}/\zeta $,  \eref{hamcavunit1} simplifies to
\begin{eqnarray}
\hat{U}(t)&=&\exp\left\{{-i[\omega_{k}t \hat{k}^{\dagger} \hat{k}-\frac{g_{k}^{2}}{\zeta^{2}}\left[\omega_{m}t-\sin( \zeta t)](\hat{k}^{\dagger}\hat{k})^{2}
+\beta t(\hat{a}^{\dagger}\hat{a})^{2}\right]}\right\}\nonumber\\
&&\times\exp\left[{\frac{g_{k}}{\zeta}\hat{k}^{\dagger}\hat{k} (\hat{a}^{\dagger}-\hat{a})-\frac{g_{k}}{\zeta}\hat{k}^{\dagger}\hat{k}(\hat{a}^{\dagger}e^{-i\zeta t}-\hat{a}e^{i\zeta t})}\right]
\exp\left[{-i\zeta t \hat{a}^{\dagger} \hat{a}}\right].\nonumber
\label{hamcavunit2}
\end{eqnarray}


\begin{thebibliography}{99}
\bibitem{nonlinbe}  A nonlinear oscillator may have an amplitude-dependent oscillation frequency and exhibit for instance quantum chaotic features. See, e.g.,  \cite{scndrf} and \cite{scndrftwo}.
\bibitem{canti}We note that our analysis could equally well be applied to a doubly clamped resonator or a membrane.
\end{thebibliography}
\end{document}